\documentclass[a4paper,usenatbib,times]{mn2e}
\usepackage{natbib,graphicx,amsmath,amsfonts,amssymb,times,txfonts,color,bm}

\def\sl{S_{\mathrm{l}}}

\def\ppd{protoplanetary disc}
\def\sopt{s_{\mathrm{opt}}}
\def\rhog{\rho_{\mathrm{g}}}
\def\rhod{\rho_{\mathrm{d}}}
\def\ts{t_{\mathrm{s}}}
\def\sspd{c_{\mathrm{s}}}
\def\vd{\textbf{v}_{\mathrm{d}}}
\def\vg{\textbf{v}_{\mathrm{g}}}

\def\vkz{v_{\mathrm{k,0}}}

\def\sz{S_{\mathrm{0}}}

\def\etaz{\eta_{\mathrm{0}}}

\def\St{\mathrm{St}}
\def\dst{\displaystyle}

\def\rhog{\rho_{\mathrm{g}}}
\def\rhod{\rho_{\mathrm{d}}}
\def\cs{c_{\mathrm{s}}}
\def\brhog{\bar{\rho}_{\mathrm{g}}}
\def\bcs{\bar{c}_\mathrm{s}}

\def\Rz{r_{0}}
\def\csz{c_{\mathrm{s}0}}

\def\Sigmaz{\Sigma_{0}}
\def\vkz{v_{\mathrm{k}0}}

\def\Hz{H_{0}}

\def\tsz{t_{\mathrm{s}0}}

\def\md{m_{\mathrm{d}}}
\def\ts{t_{\mathrm{s}}}

\def\tvr{\tilde{v}_{r}}
\def\tvtheta{\tilde{v}_{\theta}}

\def\vkz{v_{\mathrm{k}0}}
\def\phiz{\phi_{0}}
\def\etaz{\eta_{0}}
\def\etazsq{\eta_{0}^{2}}
\def\sz{S_{0}}
\def\szsq{S_{0}^{2}}

\def\Tm{T_{\mathrm{m}}}
\def\Tml{T_{\mathrm{m},\Lambda}}

\def\Rc{R_{\mathrm{c}}}

\def\Lambdaa{\Lambda_{\mathrm{A}}}
\def\Lambdab{\Lambda_{\mathrm{B}}}
\def\Lambdau{\Lambda_{\mathrm{1}}}
\def\Lambdad{\Lambda_{\mathrm{2}}}

\def\Tn{T_{\mathrm{n}}}

\def\Ru{R_{1}}
\def\Tu{T_{1}}
\def\Su{S_{1}}
\def\Td{T_{2}}
\def\Rua{R_{1}^{p+q+\frac{1}{2}}}

\newcommand{\ind}[2]{#1_\mathrm{#2}}
\newcommand{\ddt}[1]{\frac{\mathrm{d} \!\! \ #1}{\mathrm{d} t}}

\newenvironment{mylist}{\begin{list}{-}{\topsep=0pt \parskip=0pt
\itemsep=0pt \leftmargin=0.6cm}}{\end{list}}

\title[Growing dust grains: radial drift with toy models]{Growing dust grains in protoplanetary discs --- I. Radial drift with toy growth models}

\author[G. Laibe et al.]{Guillaume Laibe$^{1}$\thanks{E-mail:guillaume.laibe@monash.edu}, Jean-Fran\c cois Gonzalez$^{2}$, Sarah T. Maddison$^{3}$\\
$^{1}$Monash Centre for Astrophysics (MoCA) and School of Mathematical Sciences, Monash University, Clayton, Vic 3800, Australia\\
$^{2}$Universit\'e de Lyon, Lyon, F-69003, France; Universit\'e Lyon 1, Villeurbanne, F-69622, France; CNRS, UMR 5574, Centre de Recherche Astrophysique de Lyon;\\
 \'Ecole normale sup\'erieure de Lyon, 46, all\'ee d'Italie, F-69364 Lyon Cedex 07, France\\
$^{3}$Centre for Astrophysics and Supercomputing, Swinburne University,
PO Box 218, Hawthorn, VIC 3122, Australia
}

\pagerange{\pageref{firstpage}--\pageref{lastpage}} \pubyear{2012}

\begin{document}
%
%
%


\def\jnl@style{\it}
\def\aaref@jnl#1{{\jnl@style#1}}

\def\aaref@jnl#1{{\jnl@style#1}}

\def\aj{\aaref@jnl{AJ}}                   
\def\araa{\aaref@jnl{ARA\&A}}             
\def\apj{\aaref@jnl{ApJ}}                 
\def\apjl{\aaref@jnl{ApJ}}                
\def\apjs{\aaref@jnl{ApJS}}               
\def\ao{\aaref@jnl{Appl.~Opt.}}           
\def\apss{\aaref@jnl{Ap\&SS}}             
\def\aap{\aaref@jnl{A\&A}}                
\def\aapr{\aaref@jnl{A\&A~Rev.}}          
\def\aaps{\aaref@jnl{A\&AS}}              
\def\azh{\aaref@jnl{AZh}}                 
\def\baas{\aaref@jnl{BAAS}}               
\def\icarus{\aaref@jnl{icarus}} 
\def\jrasc{\aaref@jnl{JRASC}}             
\def\memras{\aaref@jnl{MmRAS}}            
\def\mnras{\aaref@jnl{MNRAS}}             
\def\pra{\aaref@jnl{Phys.~Rev.~A}}        
\def\prb{\aaref@jnl{Phys.~Rev.~B}}        
\def\prc{\aaref@jnl{Phys.~Rev.~C}}        
\def\prd{\aaref@jnl{Phys.~Rev.~D}}        
\def\pre{\aaref@jnl{Phys.~Rev.~E}}        
\def\prl{\aaref@jnl{Phys.~Rev.~Lett.}}    
\def\pasp{\aaref@jnl{PASP}}               
\def\pasj{\aaref@jnl{PASJ}}               
\def\qjras{\aaref@jnl{QJRAS}}             
\def\skytel{\aaref@jnl{S\&T}}             
\def\solphys{\aaref@jnl{Sol.~Phys.}}      
\def\sovast{\aaref@jnl{Soviet~Ast.}}      
\def\ssr{\aaref@jnl{Space~Sci.~Rev.}}     
\def\zap{\aaref@jnl{ZAp}}                 
\def\nat{\aaref@jnl{Nature}}              
\def\iaucirc{\aaref@jnl{IAU~Circ.}}       
\def\aplett{\aaref@jnl{Astrophys.~Lett.}} 
\def\apspr{\aaref@jnl{Astrophys.~Space~Phys.~Res.}}
\def\bain{\aaref@jnl{Bull.~Astron.~Inst.~Netherlands}} 
\def\fcp{\aaref@jnl{Fund.~Cosmic~Phys.}}  
\def\gca{\aaref@jnl{Geochim.~Cosmochim.~Acta}}   
\def\grl{\aaref@jnl{Geophys.~Res.~Lett.}} 
\def\jcp{\aaref@jnl{J.~Chem.~Phys.}}      
\def\jgr{\aaref@jnl{J.~Geophys.~Res.}}    
\def\jqsrt{\aaref@jnl{J.~Quant.~Spec.~Radiat.~Transf.}}
\def\memsai{\aaref@jnl{Mem.~Soc.~Astron.~Italiana}}
\def\nphysa{\aaref@jnl{Nucl.~Phys.~A}}   
\def\physrep{\aaref@jnl{Phys.~Rep.}}   
\def\physscr{\aaref@jnl{Phys.~Scr}}   
\def\planss{\aaref@jnl{Planet.~Space~Sci.}}   
\def\procspie{\aaref@jnl{Proc.~SPIE}}   

\let\astap=\aap
\let\apjlett=\apjl
\let\apjsupp=\apjs
\let\applopt=\ao

\label{firstpage}
\bibliographystyle{mn2e}
\maketitle

\begin{abstract}

In a series of papers, we present a comprehensive analytic study of the global motion of growing dust grains in protoplanetary discs, addressing both the radial drift and the vertical settling of the particles. Here we study how the radial drift of dust particles is affected by grain growth. In a first step, toy models in which grain growth can either be constant, accelerate or decelerate are introduced. The equations of motion are analytically integrable and therefore the grains dynamics is easy to understand.

The radial motion of growing grains is governed by the relative efficiency of the growth and migration processes which is expressed by the dimensionless parameter $\Lambda$, as well as the exponents for the gas surface density and temperature profiles, denoted $p$ and $q$ respectively. When $\Lambda$ is of order unity, growth and migration are strongly coupled, providing the most efficient radial drift. For the toy models considered, grains pile up when $-p+q+1/2<0$. Importantly, we show the existence of a second process which can help discs to retain their solid materials. For accelerating growth, grains end up their migration at a finite radius, thus avoiding being accreted onto the central star.

\end{abstract}


\section{Introduction}
\label{sec:intro}

Various mechanisms have been suggested to explain the formation of hundred metre-sized bodies called planetesimals found in protoplanetary discs. \citet{GoldreichWard1973} proposed that planetesimals could form directly by gravitational collapse  when dust settles and concentrates in the midplane. While this leads to rapid planetesimal formation, the process requires a large dust-to-gas ratio, which is prevented by the Kelvin-Helmholtz instability \citep{Cuzzi1993,GaraudLin2004,Chiang2008,Barranco2009,Lee2010a,Lee2010b}. Other processes have therefore been suggested, such as the streaming instability due to the collective effects of dust \citep{Goodman2000,Youdin2005,Bai2010a,Bai2010b,Jacquet2011}, followed by gravitational instability \citep{Johansen2007} or turbulent concentration between eddies \citep{Cuzzi2008,Shi2013}. Whatever the mechanism, planetesimal formation requires the presence of smaller primitive bodies called pre-planetesimals. The existence of our own solar system and of extrasolar planets, now commonly observed, obviously implies that at least in some protoplanetary discs pre-planetesimals can form.

Pre-planetesimals are thought to form by grain coagulation: sub-micron sized dust particles which originated in the interstellar medium or condensed out of the newly formed nebula collide and stick under the influence of the Van der Waals interaction \citep[e.g.][]{Chokshi1993,Cuzzi1993,Blumwrum2008}. Recent observations of protoplanetary discs \citep{Testi2003,Apai2005,Lommen2007,Ricci2010} confirm the presence of small growing grains which can reach at least centimetre sizes \citep[e.g.][]{Wilner2003,Lommen2009}. It has been difficult, however, to explain how grains could form bodies larger than a typical size of ten centimetres without bouncing or fragmenting \citep{Blumwrum2008,ChiangYoudin2010}. Recently, a novel scenario which includes a realistic probability distribution function for the dust velocities overcomes this issue \citep{Garaud2013}. However, the question of the nature of the global evolution of grains in protoplanetary discs remains open.

\defcitealias{Weidendust1977}{W77}\defcitealias{Nakagawa1986}{NSH86}\defcitealias{YS2002}{YS02}\defcitealias{Laibe2012}{LGM12}
Although the dynamics of non-growing dust grains in discs has been well studied \citep[hereafter W77, NSH86,YS02 and LGM12]{Weidendust1977,Nakagawa1986,YS2002,Laibe2012}, numerical studies describing the evolution of growing grains are more recent \citep{Schmitt1997, StepVal1997, Suttner1999, Tanaka2005, DullemondDom2005, Klahr2006, Garaud2007, Brauer2008, Laibe2008, Birnstiel2009,Birnstiel2010a}. Generally, for reasons of computational cost, a compromise between degrees of refinement of the disc hydrodynamics and the grain growth treatment is required: 2D or 3D studies usually have a rather crude growth model whereas works with a more sophisticated growth treatment are generally only 1D. The main result of these studies is that grain growth occurs very quickly (micron-sized particles reach centimetre sizes in a few 1000~yr). Such behaviour implies that the formation of millimetre-sized bodies in {\ppd}s is very easy. Importantly, \citet{Brauer2008} overturned the commonly held idea that all the dust particles are accreted onto the central star once they reach a size for which the radial migration velocity is largest (the so-called ``radial-drift barrier''). Indeed, if the growth timescale is smaller than the migration timescale of the grains, the growth is efficient enough for the grains to decouple from the gas before being accreted by the central star.

In our recent study \citep{Laibe2012}, we revisited the classical theory of the so-called ``radial-drift barrier'',  highlighting the role of both the drag-dominated (or A-mode) and the gravity dominated (or B-mode) of migration and showing that the grain pile-up efficiently prevents accretion of grains on the host star where the disc structure of observed discs is considered. YS02 already found this result for the A-mode ; we generalised it to the B-mode, motivated by the importance of this regime when grains are growing. The results of this study therefore suggest that the radial-drift barrier may not be systematic for growing grains as well. However, the existing analytical derivations of growing grains --- e.g. in \citet{Garaud2007}, \citet{Brauer2008}, and \citet{Laibe2008} ---  are not refined enough to quantitatively constrain which discs are able to retain their grains and allow planetesimals to grow.

In a series of papers (hereafter Papers~I, II, III), we study analytically the dynamics of growing grains, namely their radial drift and vertical settling. In this paper (Paper~I), we focus on the radial motion of the grains, starting with toy models of grain growth. The results and the concepts introduced during this study are indeed of great help for understanding physical growth models. We introduce the equations of motion in Sect.~\ref{sec:nut} and consider the toy growth models in Sect.~\ref{sec:growth}. We derive the radial evolution in the case of linear grain growth in Sect.~\ref{sec:AnalyticalLinear} where we explain some fundamental features of the grain evolution. Finally, we highlight the importance of accelerating growth rates on the dust radial motion in Sect.~\ref{sec:GeneralCase}, studying growth as a power law. Paper~II extends this study to the case of realistic growth models and Paper~III deals with the vertical settling of growing grains.

\section{Radial drift of non-growing grains}
\label{sec:nut}

\subsection{Equations of motion}

Let us first consider some important properties of the dynamics of a non-growing grain in a laminar non-magnetic and non self graviting protoplanetary discs (the study of a turbulent disc is beyond the scope of this paper). Grain dynamics is dominated by two forces: the central star's gravity, which tends to make grains orbit in a Keplerian manner, and interactions with gas particles, which are macroscopically represented by a drag force. If a dust grain has a velocity which is different to that of the gas, it experiences a drag force which tends to damp this differential velocity. The expression of the gas drag in the Epstein regime is given by:
\begin{equation}
\left\lbrace
\begin{array}{rcl}
\mathbf{F}_\mathrm{d} & = & \dst\frac{m_\mathrm{d}}{\ts}\,\Delta\mathbf{v} \, , \\[2ex]
\ts & = & \dst\frac{\rhod s}{\rhog \sspd} \, ,
\end{array}
\right. 
\label{eq:Epsts}
\end{equation}
where $m_\mathrm{d}$ is the dust grain's mass, $\ts$ the stopping time, $\rhog$ the gas density, $\sspd$ the local gas sound speed, $\rhod$ the intrinsic dust density, $s$ the grain size, and $\Delta\mathbf{v} = \ind{\mathbf{v}}{d} - \ind{\mathbf{v}}{g}$ the differential velocity between dust and the mean gas motion. In this study, we do not consider the Stokes drag. Indeed, when solid bodies are large enough to migrate in this friction regime, they are not expected to grow by hit-and-stick collisions. However, the mathematical formalism remains the same and the results can be easily translated from one regime to he other one as in  \citetalias{Laibe2012}.

Seminal studies of dust dynamics were conducted by \citet{Whipple1972}, \citetalias{Weidendust1977}, \citet{Weiden1980} and \citetalias{Nakagawa1986}, and extended by others \citep{Takeuchi2002,Haghighipour2003,Garaud2004,YoudinChiang2004}. Here we recall the major points of those studies. The equation of motion of the grain is given by
\begin{equation}
\ddt{\vd} + \frac{1}{\ts} \left( \vd - \vg \right) +  \textbf{g} = 0 \, .
\label{eq:equagrain}
\end{equation}
We now use the notations and the dimensionless quantities introduced in \citetalias{Laibe2012} and given in Appendix~\ref{App:Notations}. The radial surface density $\Sigma$ and temperature $\mathcal{T}$ profiles are described by power-laws $\Sigma \propto R^{-p}$ and $\mathcal{T} \propto R^{-q}$.  the physical relations are written in cylindrical coordinates ($r$, $\theta$, $z$). The equation of motion of such grains are given by (see  \citetalias{Laibe2012})
\begin{equation}
\left\lbrace 
\begin{array}{rcl}
\dst \frac{\mathrm{d} \tvr}{\mathrm{d} T} - \frac{\tvtheta^{2}}{R} + \frac{\tvr}{\sz}R^{-\left(p+\frac{3}{2} \right)}e^{-\frac{Z^{2}}{2R^{3-q}}} +\frac{R}{\left(R^2 + \phiz Z^{2} \right)^{3/2}} & = & 0 \\[3ex]
\dst \frac{\mathrm{d} \tvtheta}{\mathrm{d} T} + \frac{\tvtheta\tvr}{R} + \frac{\tvtheta - \sqrt{\frac{1}{R} - \etaz R^{-q} - q\left(\frac{1}{R} - \frac{1}{\sqrt{R^{2}+\phiz Z^{2}}} \right)}}{\sz}\\
\dst\times\ R^{-\left(p+\frac{3}{2} \right)}e^{-\frac{Z^{2}}{2R^{3-q}}} & = & 0 \\[2ex]
\dst \frac{\mathrm{d}^{2}Z}{\mathrm{d}T^{2}} + \frac{1}{\sz}\frac{\mathrm{d}Z}{\mathrm{d}T}R^{-\left(p+\frac{3}{2} \right)}e^{-\frac{Z^{2}}{2R^{3-q}}} + \frac{Z}{\left(R^{2} + \phiz Z^{2} \right)^{3/2}} & = & 0 ,
\end{array}
\right.
\label{eq:dust3d}
\end{equation}
where $T$, $R$, $Z$, $\sz$ are the dimensionless time, radius, height and initial grain size respectively These equations depend on five parameters ($\etaz$, $\sz$, $\phiz$, $p$, $q$) which are the initial dimensionless acceleration due to the pressure gradient, initial grain size, initial disc aspect ratio and exponents of surface density and temperature profiles.

\subsection{The Stokes number}

The ratio $\frac{\ts}{t_{\mathrm{k}}}$ of the drag and the Keplerian timescales is a dimensionless parameter of the problem often called the Stokes number and frequently denoted by $\St$:
\begin{equation}
\St = \frac{\ts}{t_{\mathrm{k}}} = 
\frac{s}{\frac{\rhog \cs}{\rhod \Omega_{\mathrm{k}}}} = \frac{s}{s_{\mathrm{opt}}} =
 \sz R^{p}e^{\frac{Z^{2}}{2R^{3-q}}},
\label{eq:defsz}
\end{equation}
where $ \Omega_{\mathrm{k}}$ is the Keplerian frequency. The size $s_{\rm opt}$, which we call the optimal size for radial migration \citep[see][and Sect.~\ref{sec:nut}]{Fouchet2007}, is of the order of 1~m in a Minimum Mass Solar Nebula (MMSN) and is related to the so-called ``metre-size barrier'', first mentioned in \citetalias{Weidendust1977}, which is thought to prevent pre-planetesimal formation. However, the value of $\sopt$ varies with $r$ and depends strongly on the disc structure, which for observed discs is very different from that of the MMSN, explaining why the term ``radial-drift barrier'' is more appropriate.

\subsection{Radial drift}

When $Z = 0$, this implies
\begin{equation}
\St = \sz R^{p} .
\label{eq:swg}
\end{equation}
The radial dust motion is therefore governed by the equations
\begin{equation}
\left\lbrace
\begin{array}{l}
\dst \frac{\mathrm{d} \tvr}{\mathrm{d} T}  =  \dst \frac{\tvtheta^{2}}{R} - \frac{\tvr}{\sz}R^{-\left(p+\frac{3}{2} \right)} -\frac{1}{R^2} \\[2ex]
\dst \frac{\mathrm{d} \tvtheta}{\mathrm{d} T}  =  \dst -\frac{\tvtheta \tvr}{R} - \frac{\tvtheta - \sqrt{\frac{1}{R} - \etaz R^{-q}}}{\sz}R^{-\left(p+\frac{3}{2} \right)} .
\end{array}
\right.
\label{eq:radialseulmodif}
\end{equation}
This radial motion of dust particles is a well studied problem \citepalias{Weidendust1977}. Because of the radial pressure gradient which counters the gravity of the central star, the azimuthal velocity of the gas is sub-Keplerian. The sub-Keplerian gas motion is the source of the differential velocity between gas and dust, which tends to be attenuated by gas drag and leads to a radial drift motion or migration of the dust (it  should not be confused with planetary migration, which has a different physical origin, namely tidal torques). Therefore, two effects control the radial motion: gas drag (whose intensity depends on $\St$) and the sub-Keplerian motion of the gas, which produces the differential velocity and is characterized by the $ \etaz $ parameter. An expansion at small pressure gradients of order $\mathcal{O}\left(\etaz \right)$ (see \citetalias{Nakagawa1986}), reduces Eq.~\ref{eq:radialseulmodif} to
\begin{equation}
\tvr = -\frac{\etaz \sz R^{p-q+\frac{1}{2}} }{1 + R^{2p} \szsq} + \mathcal{O}\left(\etazsq \right). 
\label{eq:nakatvrusimp}
\end{equation}
Thus, depending on the value of $\St$, the particle experiences various modes of inward drift, as stressed in \citetalias{Laibe2012}.

If $\St = \sz R^{p} \ll 1$ (which we call the A-mode, where drag dominates over gravity), particles are damped in a few stopping times, which is smaller than the orbital timescale. The dust is then forced by the gas to orbit at a sub-Keplerian velocity, and hence can not counter the gravity of the central star and migrates inward. This radial motion is then quickly counterbalanced by gas damping: particles migrate with a limited radial drift velocity, which is low for the smallest grains for whom the drag is strongest. If such particles have an initial radial velocity, drag circularizes their orbit in a few $ \ts $. In the A-mode, Eq.~\ref{eq:nakatvrusimp} reduces to
\begin{equation}
\tvr = \frac{\mathrm{d} R}{\mathrm{d} T} = -\etaz \sz R^{p-q+\frac{1}{2}} .
\label{eq:radmodeA}
\end{equation}

When $\St = \sz R^{p} \gg 1$ (which we call the B-mode, where gravity dominates over drag), particles have a stopping time larger than the orbital timescale: they orbit with an approximately Keplerian velocity. Gas drag exerts a torque which makes the particles lose angular momentum. As the angular momentum is an increasing function of the radius ($l \propto \sqrt{r}$), the dust migrates inwards. The larger the particle, the longer the response to the torque action and thus larger particles migrate at a slower rate. However, $ \mathrm{d}l/\mathrm{d}r \propto r^{-1/2}$ is a decreasing function of radius. To lose the same amount of angular momentum, grains in the inner part of the disc cover less distance than grains in the outer part: the radial drift motion becomes less efficient as dust migrates closer to the central star. In the B-mode, Eq.~\ref{eq:nakatvrusimp} reduces to
\begin{equation}
\tvr = \frac{\mathrm{d} R}{\mathrm{d} T} = -\frac{\etaz}{\sz} R^{-p-q+\frac{1}{2}}.
\label{eq:radmodeB}
\end{equation}

\begin{figure}
\resizebox{\hsize}{!}{\includegraphics[angle=-90]{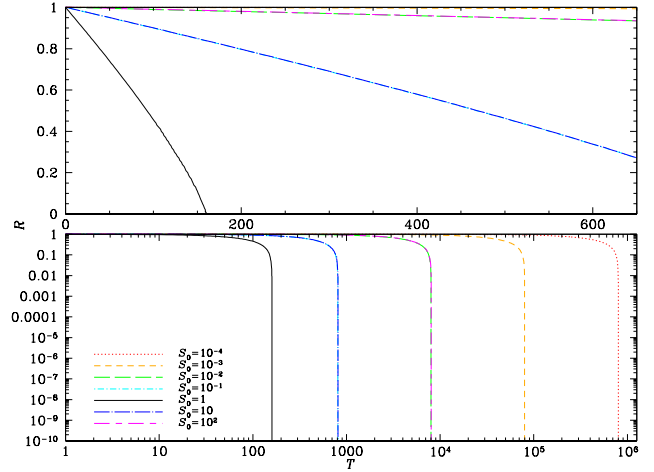}}
\caption{Radial motion $R\left(T\right)$ of non-growing dust grains for $\etaz = 10^{-2}$, $p=0$ and $q=3/4$ ($-p+q+\frac{1}{2}>0$). $\sz$ varies from $10^{-4}$ to $10^{2}$. Top: linear scale, bottom: logarithmic scale. Grains are accreted in a finite time onto the central star.} 
\label{fig:plotfirstcase}
\end{figure}

\begin{figure}
\resizebox{\hsize}{!}{\includegraphics[angle=-90]{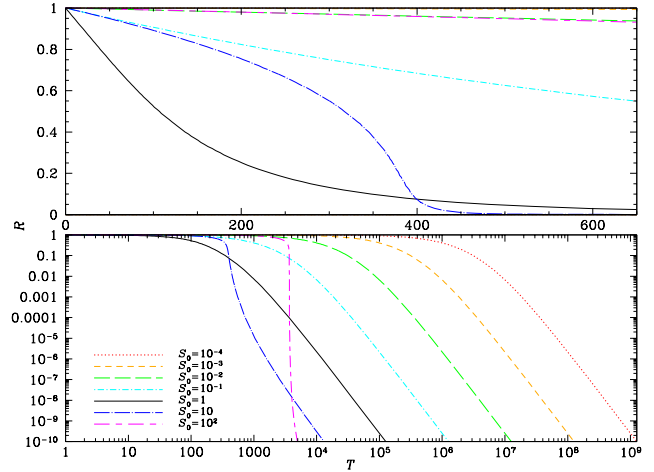}}
\caption{Same as Fig.~\ref{fig:plotfirstcase} for $p=3/2$ and $q=3/4$ ($-p+q+\frac{1}{2}<0$). Grains experience a pile-up which prevent them to be accreted onto the central star. Compare the black curve of this figure with the black curve of Fig.~\ref{fig:plotfirstcase} to clearly see the grain's pile-up.} 
\label{fig:plotsecondcase}
\end{figure}

If $\St \thicksim 1$, both previous effects combine, leading to a large radial drift velocity: particles with such sizes migrate very efficiently onto the central star.

However, since $\St = S_{0}R^{p}$ tends to zero at small radii, the grains radial motion ends in the A-mode. Integrating Eq.~8 gives the time at which a migrating grains reaches
a given radius $R$:
\begin{equation}
T(R)=\frac{1 - R^{-p+q+\frac{1}{2}}}{\left(-p+q+\frac{1}{2}\right)\eta_0 S_0}.
\label{eq:tgene}
\end{equation}
if $-p+q+\frac{1}{2}\ne0$ (and $T=-\mathrm{ln}(R)/(\eta_0 S_0)$ if
$-p+q+\frac{1}{2}=0$). The final outcome of the grains is therefore determined
by the sign of $-p+q+\frac{1}{2}$ (LGM12). This criterion determines
the effect of the acceleration due to the increase of the pressure gradient
which is counterbalanced by the increase of gas drag. If $-p+q+\frac{1}{2}>0$,
the dust particles are accreted onto the central star in a finite time
\begin{equation}
T(R=0)=\frac{1}{\eta_0 S_0 \left(-p+q+\frac{1}{2}\right)},
\label{eq:tacc}
\end{equation}
 but if $-p+q+\frac{1}{2}\le0$
\begin{equation}
T(R)=\mathcal{O}\left(R^{-p+q+\frac{1}{2}}\right)
\label{eq:tpileup}
\end{equation}
for small values of $R$ and the dust particles take an infinite time to be accreted
(i.e. grains pile up and survive the radial-drift barrier), or more rigorously,
the time required to reach $R\to0$ diverges as a power law. Importantly, this also means that, although the grain's kinematics
depends on the local values of the velocities, its outcome results from the
integration of its global motion through the disc. Figs.~\ref{fig:plotfirstcase} and \ref{fig:plotsecondcase} show plots of the dust radial motion for different sizes $\sz$ and different signs of $-p+q+\frac{1}{2}$ For the remainder of this paper, we study grain growth and will present similar plots to Figs.~\ref{fig:plotfirstcase} and \ref{fig:plotsecondcase}, allowing easy comparison of the effects grain growth has on the dynamics. In  \citetalias{Laibe2012}, we have shown that finite disc inner radii and disc lifetime only slightly mitigate this result: in a Classical T-Tauri Star disc, the grains pile-up prevents almost all the grains up to millimetre sizes in the disc from being accreted.

\section{Toy models of  grain growth}
\label{sec:growth}

\subsection{The Stokes number}
\label{sec:stgrowth}
To study grain growth, we generalise the relation of Eq.~\ref{eq:swg} by introducing the quantity $S(T)$ defined by
\begin{equation}
\St = S R^{p} .
\label{eq:defsbarnew}
\end{equation}
where $S$ corresponds to the part of the Stokes number which is related only to the grain size and does not depend on the radial coordinate. $\St$ and $S$ are related to the physical quantities $s$, $s_{\rm opt}$ and $s_{\rm opt,0}$ via:
\begin{equation}
\St = \frac{s}{s_{\rm opt}},
\end{equation}
and 
\begin{equation}
S = \frac{s}{s_{\rm opt, 0}} .
\end{equation}
The initial grain size is denoted $s(t=0) = s_{0}$ and we define
\begin{equation}
\sz \equiv \frac{s_{0}}{s_{\rm opt, 0}} = S(T = 0) .
\end{equation}
Since initially $R=1$, we have also
\begin{equation}
\St(T = 0) = \sz.
\end{equation}
From a practical point of view, all the equations of motion derived from the NSH86 expansion in \citetalias{Laibe2012} can be used for studying the grain growth, replacing $\sz$ by $S$ since they do not depend on the grain size. The evolution of $S(T)$ has to be itself specified by a model of grain growth.
 
\subsection{Equations of evolution}

Retaining full generality, the equations of motion given by Eq.~\ref{eq:radialseulmodif} become
\begin{equation}
\left\lbrace
\begin{array}{l}
\dst \frac{\mathrm{d} \tvr}{\mathrm{d} T}  =  \dst \frac{\tvtheta^{2}}{R} - \frac{\tvr}{S\!\left( T \right)}R^{-\left(p+\frac{3}{2} \right)} -\frac{1}{R^2} \\[3ex]
\dst \frac{\mathrm{d} \tvtheta}{\mathrm{d} T}  =  \dst -\frac{\tvtheta \tvr}{R} - \frac{\tvtheta - \sqrt{\frac{1}{R} - \etaz R^{-q}}}{S\!\left( T \right)}R^{-\left(p+\frac{3}{2} \right)} .
\end{array}
\right.
\label{eq:eqradwithgrowth}
\end{equation}
The expansion with respect to $\etaz$ does not depend on $S(T)$ and can be generalized when the grain size depends on time. Thus, substituting Eq.~\ref{eq:defsbarnew} into Eq.~\ref{eq:nakatvrusimp}, the expression of $\frac{\mathrm{d} R}{\mathrm{d} T}$ becomes
\begin{equation}
\tvr = \frac{\mathrm{d} R}{\mathrm{d} T} = \frac{-\etaz S\!\left( T \right) R^{p-q+\frac{1}{2}}}{1 + R^{2p}S\!\left( T \right)^{2n}} + \mathcal{O}(\etaz^{2}).
\label{eq:NGS86withgrowth}
\end{equation}

The remaining equation describes the evolution of $S(T)$. This can be rigorously determined from the physical growth rates (see Paper~II) or prescribed by a toy model.
 
\subsection{Toy growth models}

We start to study the radial motion of growing grains using toy growth models for two reasons. First, we want to understand the impact of the grain size evolution on the dynamics independently of the radial dependence of the physical growth rate. We shall see that even with a  prescribed growth rate, the dust behaviour may be rather complex. Second, the equation of evolution are analytically tractable. It should be noted that the toy models we shall use correspond for some specific configurations to physical growth rates (see Paper~II).

The simplest toy model consists of the following linear relation:
\begin{equation}
S = S_{0} + \gamma T .
\label{eq:toy_lin}
\end{equation}
Eq.~\ref{eq:toy_lin} is equivalent to setting:
 \begin{equation}
 s(t) = s_{0} + \dot{s}_{0} t ,
 \end{equation}
giving
\begin{equation}
\gamma = \frac{\dot{s}_{0}}{s_{\rm opt, 0}/t_{\rm k,0}} 
\end{equation}
for the dimensionless parameter which measures the growth efficiency. Even though this model is the simplest to treat the grain growth, it can not handle the cases of accelerating or decelerating growth. To be able to investigate what happens in those cases, we adopt a power law prescription for the grain sizes given by: 
\begin{equation}
S\left(T\right) = \left(S_{0}^{1/n} + \gamma T \right)^{n} .
\label{eq:powergrowth}
\end{equation}
The initial size $\sz$ is chosen to be small enough  (rigorously, $\sz \ll \mathcal{O}(1)$) so that grains will experience all drag regimes as they grow. Indeed, varying $\gamma$ modifies the growth rate and choosing $n<1$, $n>1$ or $n=1$ ensures the growth slows down, speeds up or keeps a constant rate, respectively. While crude, these models allow us firstly to separate the effect coming from the increase of the grain size only to that due to the radial dependancy of the growth model, and secondly to distinguish different dynamical evolutions for the growing grains by allowing an analytical integration of the equations of motion. Since we are considering grain growth, $ \gamma > 0 $.

\section{The linear growth regime}
\label{sec:AnalyticalLinear}

\subsection{Analytic solution}

For the linear growth case, we give an analytical expression for the grain size evolution which provides important information on the radial behaviour of grains. Starting from
\begin{equation}
\frac{\mathrm{d} R}{\mathrm{d} T} = \frac{-\etaz \left(S_{0} + \gamma T \right) R^{p-q+\frac{1}{2}}}{1 + R^{2p} \left(S_{0} + \gamma T \right)^{2}}
\label{eq:linavantchange}
\end{equation}
and setting $u = \left(S_{0} + \gamma T \right)^{2}$ gives
\begin{equation}
\frac{\mathrm{d} R}{\mathrm{d} u} = \frac{\mathrm{d} R}{\mathrm{d} T} \frac{\mathrm{d} T}{\mathrm{d} u} = - \frac{\etaz}{2 \gamma} \frac{R^{p-q+\frac{1}{2}}}{1+R^{2p} u} .
\label{eq:linchangevar}
\end{equation}
Here, the dimensionless parameter, $\Lambda$ defined by
\begin{equation}
\Lambda = \frac{\gamma}{\etaz} ,
\label{eq:deflamda}
\end{equation}
naturally arises from the equations. It compares the efficiency of the growth and the migration processes. Eq.~\ref{eq:linchangevar} can be rewritten as
\begin{equation}
\frac{\mathrm{d} u}{\mathrm{d} R} + 2 \Lambda R^{p+q-\frac{1}{2}} u  = - 2 \Lambda R^{-p+q-\frac{1}{2}} ,
\label{eq:eqdifflin}
\end{equation}
which can be integrated by the method of variation of parameters. Assuming that both $p$ and $q$ take positive values as expected for real discs, $p+q+\frac{1}{2} > 0$ and we obtain the following expression
\begin{equation}
\begin{array}{rcl}
T(R) & = & \dst\frac{1}{\gamma}\left[\sqrt{u(R)}-\sz\right]\\[2ex]
&=& \dst\frac{1}{\gamma} \left[ \sqrt{S_{0}^{2} e^{2 \Lambda \frac{1 - R^{p+q+\frac{1}{2}}}{p+q+\frac{1}{2}}} +  2\Lambda e^{-2 \Lambda \frac{R^{p+q+\frac{1}{2}}}{p+q+\frac{1}{2}}}J(R)} - S_{0} \right] ,
\end{array}
\label{eq:exprintlin}
\end{equation}
where
\begin{equation}
J\left(R\right) = \int_{R}^{1} \hat{R}^{-p+q-\frac{1}{2}} e^{2 \Lambda \frac{\hat{R}^{p+q+\frac{1}{2}}}{p+q+\frac{1}{2} }} \mathrm{d}\hat{R}
 = \frac{1}{x} \int_{w^{1/x}}^{1} \hat{w}^{\frac{y+1}{x}-1} e^{\frac{2\Lambda}{x}\hat{w}} \mathrm{d}\hat{w},
\label{eq:defJ}
\end{equation}
with
\begin{equation}
\left\lbrace
\begin{array}{rcl}
x & = & p+q+\frac{1}{2} > 0\\ [2 ex]
y & = & -p+q-\frac{1}{2}\\ [2 ex]
w & = & R^{p+q+\frac{1}{2}}  .
\end{array}
\right.
\label{eq:changevarint}
\end{equation}
While not transparent, the global behaviour of the radial motion of a growing dust grain strongly depends on $-p+q+\frac{1}{2}$ and $\Lambda$. The following considerations apply:
\begin{figure}
\resizebox{\hsize}{!}{\includegraphics[angle=-90]{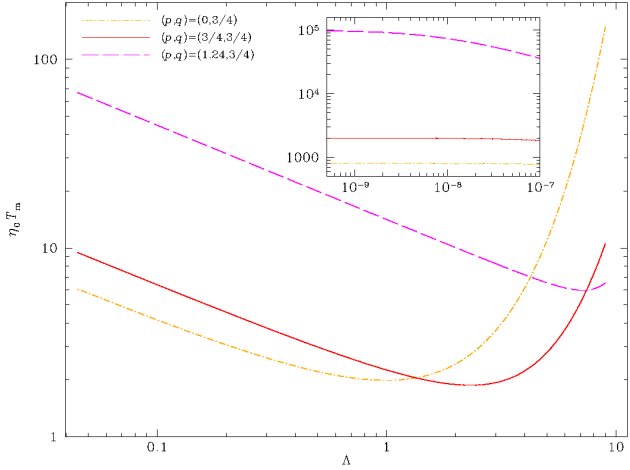}}
\caption{Accretion time as a function of $\Lambda$, the relative efficiency of growth vs. migration. Logarithmic plot of $\etaz\, \Tm \left(\Lambda \right)$ for $\sz=\mathbf{10^{-3}}$, $(p,q)=(0,3/4)$, $(3/4,3/4)$ and $(1.24,3/4)$. The plot shows an optimal accretion time for $\Lambda = \mathcal{O}(1)$ (the migration time and the growth time have the same order of magnitude) and $-p+q+1/2 > 0$ (no pile-up). The inset shows that $\lim_{\Lambda \to 0} \eta_0 T_\mathrm{m}$ is indeed finite, as shown by Eqs~\ref{eq:tmdeltap} and \ref{smalldeltap}.} 
\label{fig:plotKummer}
\end{figure}
\begin{mylist}

\item If $-p+q+\frac{1}{2} \le 0$, $J(R)$ diverges for $R \to 0$. This is the generalisation for growing grains of the result found for the radial motion of non-growing grains: the slowdown due to the gas drag dominates the acceleration due to the pressure gradient, and grains fall onto the star in an infinite time.

\item If $-p+q+\frac{1}{2} > 0$, the limit at $R \to 0$ is defined as $J(R)$ converges and the grains are accreted in a finite time $\Tm = T\left(R = 0\right)$. Taking the limit of Eq.~\ref{eq:defJ} at $R \to 0$ with $x>0$ provides
\begin{equation}
\begin{array}{rcl}
J\left(0\right) & = & \dst\frac{1}{x} \int_{0}^{1} \hat{w}^{\frac{y+1}{x}-1} e^{\frac{2\Lambda}{x}\hat{w}} \mathrm{d}\hat{w} \\[2ex]
& = & \dst\frac{1}{y+1} M\left(\frac{y+1}{x},\frac{y+1}{x}+1,\frac{2\Lambda}{x} \right),
\end{array}
\label{eq:limzeroI}
\end{equation}
where $M\left(a,b,z\right)$ is the M-Kummer confluent hypergeometric function of indices $a$ and $b$ with respect to $z$.
Thus, for $-p+q+\frac{1}{2} > 0$, the $M$ function is defined and we obtain for different values of $\gamma$ the expression of the accretion time onto the central star. Introducing $\dst a = \frac{y+1}{x} = \frac{-p+q+\frac{1}{2}}{p+q+\frac{1}{2}}$, we find the expression of the accretion time given by
\begin{equation}
\Tm = \frac{1}{\etaz\Lambda}\left[\sqrt{\szsq e^{\frac{2\Lambda}{x}} +  \frac{2\Lambda}{a x} M \left(a,a+1,\frac{2\Lambda}{x} \right) } - \sz\right] .
\label{eq:tmdeltap}
\end{equation}
\end{mylist}

\subsection{Grain radial motion}

The function $\etaz \, \Tm\left(\Lambda\right)$ is plotted in Fig.~\ref{fig:plotKummer} for $\sz=10^{-2}$ and different pairs of $(p,q)$ values. The main remarks on the shape of the curves are the following. We first fix the value of $-p+q+\frac{1}{2}>0$ and vary $\Lambda$. Noting that $M \left(a,a+1,\frac{2\Lambda}{x} \right) \to 1$ when $\Lambda \to 0$, a first order Taylor expansion of Eq. (\ref{eq:tmdeltap}) in $\Lambda$ gives
\begin{equation}
\Tm = \frac{1}{\etaz \sz} \left(\frac{\szsq}{x} + \frac{1}{a x} \right) + \frac{4\Lambda}{\etaz x^{2}} f_{a}\left(\sz \right)  + \mathcal{O}\left(\Lambda^{2} \right),
\label{smalldeltap}
\end{equation}
where
\begin{equation}
f_{a}\left(\sz \right) = \frac{1}{4}\frac{\sz^{2}\left(a + 1 \right) + 2}{\sz^{2}\left(a+1 \right)} - \frac{1}{8}\frac{\left(\sz^{2}a + 1 \right )^{2}}{a^{2}\sz^{4}} .
\end{equation}
With $x = p+q+\frac{1}{2}$ and $ax = y+1 = -p+q+\frac{1}{2}$, we recover the limit found without grain growth \citepalias[see Eq.~34 of][]{Laibe2012}. Indeed, the case $\Lambda \ll 1$ corresponds to a growth timescale that is very long compared to the migration process. Moreover, $f_{a}\left(\sz \right) = - \frac{1}{8a^{2}\sz^{4}} + \mathcal{O}\left(\sz^{-2} \right)$ when $\sz \to 0$, $f_{a}\left(\sz \right) =  \frac{1}{8} + \mathcal{O}\left(\sz^{-2} \right)$ when $\sz \to \infty$ and $f_{a}\left(\sz \right) = 0$ for $\sz  = \sqrt{\frac{ -a + 1 + \sqrt{2}\sqrt{a^{2} +1} }{ a\left(a + 1 \right) }}$. Thus, if the initial grain size is small enough, a small amount of growth decreases the accretion time. Indeed, growth takes the grain closer to the fast intermediate-size ($\St \simeq 1$) regime of migration. On the contrary, the size of an initially large grain can only grow away from the intermediate-size regime and growth slows down the dust radial motion.

Moreover, when $\Lambda \to \infty$, $\Tm \to \infty$ due to the exponential term in the square root.  More precisely,
\begin{equation}
\Tm = \frac{e^{\frac{\Lambda}{x}} \sqrt{1 + \sz^{2}} }{\etaz \Lambda} + \mathcal{O}\left(\frac{e^{\frac{\Lambda}{x}}}{\Lambda^{2}} \right).
\end{equation}
The deceleration of the migration due to the grain growth is very efficient as $\Tm$ reaches exponentially large values as $\Lambda$ increases. In this case, growth is so efficient that grains only very briefly experience the $\St = \mathcal{O}\left(1\right)$ regime of fast radial migration and are accreted onto the central star at very large times. As a result, for initially small grains, $\Tm$ reaches a minimum value as $\Lambda$ varies. The migration process is therefore optimized for a given value of $\Lambda = \frac{\gamma}{\etaz}$, when the grain size is optimized for the migration process by the growth during the whole radial motion. This minimum is called $\Tml$.

Now, if $-p+q+\frac{1}{2} \to 0$, the grain motion reaches the regime of accretion in an infinite time. As an example, the case ($p = 1.24$, $q = 3/4$) in Fig.~\ref{fig:plotKummer}, which corresponds to $-p+q+\frac{1}{2} =0.01$, shows a large value of $\Tml$ compared to the case ($p = 0$, $q = 3/4$).  If $-p+q+\frac{1}{2}$ is far enough from zero, $\Tml = \mathcal{O}\left(1\right)$ is reached for $\Lambda = \mathcal{O}\left(1\right)$. As an example, for ($p = 0$, $q = 3/4$), $\Tml = 1.98$ for $\Lambda = 1$.

\section{Growth as a power law} 
\label{sec:GeneralCase}

\subsection{Qualitative behaviour}
\begin{figure}
\resizebox{\hsize}{!}{\includegraphics{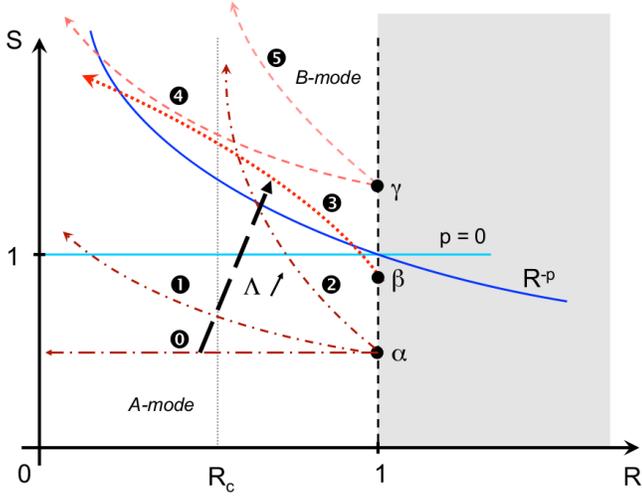}}
\caption{Trajectories of growing dust grains in the dimensionless size vs. dimensionless radius plane ($R,S$) for three initial positions $\alpha, \beta$ and $\gamma$, which are discussed in the text. Very different behaviours are expected for these grains.} 
\label{fig:Modesgrowth}
\end{figure}

We now refine the growth prescription and study the radial motion of a dust grain whose size evolution is given by the power law of Eq.~\ref{eq:powergrowth}. As for the case without growth (see Sect.~\ref{sec:nut}), information about the dust behaviour can be obtained by studying the limiting cases of the A- and B-modes. As both $S$ and $R$ vary with respect to $T$, the related mode of migration is given by:
\begin{equation}
\St(T)=\left[ S R^p\right](T).
\label{eq:defN}
\end{equation}
If $\St \ll 1$ (resp. $\St \gg 1$), grains are in the A-mode (resp. the B-mode). Thus, the evolution of $R\left(T\right)$ for grains in the A-mode of migration satisfies (see Eq.~\ref{eq:radmodeA})
\begin{equation}
\frac{\mathrm{d} R}{\mathrm{d} T} = - \etaz \left(S_{0}^{1/n} + \gamma T \right)^{n} R^{p - q + \frac{1}{2}} ,
\label{eq:Amodgrowth}
\end{equation}
and if the grain migrates in the B-mode (see Eq.~\ref{eq:radmodeB})
\begin{equation}
\frac{\mathrm{d} R}{\mathrm{d} T} = -\frac{\etaz}{\left(S_{0}^{1/n} + \gamma T \right)^{n}} R^{-p-q+\frac{1}{2}} .
\label{eq:Bmodegrowth}
\end{equation}
For clarity, we are now describing the grain motion in the $(R,S)$ plane, non-growing grains move horizontally from right to left. The drag-dominated A-mode lies below the $R^{-p}$ curve, while the gravity dominated B-mode lies above the $R^{-p}$ curve. As $S$ increases with time $T$, the grain does not move horizontally in the $(R,S)$ plane anymore. Consequently, the dynamics of the grain has some additional properties compared to the case without growth. Qualitatively, Fig.~\ref{fig:Modesgrowth} shows various outcomes for the growing grain radial motion depending on values of $\Lambda$, $n$ and on the initial grain size, represented by initial positions $\alpha$, $\beta$ (A-mode) and $\gamma$ (B-mode) in the $(R,S)$ plane:

\begin{itemize}
\item Case 0 ($\Lambda = 0$, initial position $\alpha$): Grain migrate without growing, stay in the A-mode and reach the central star in a finite or infinite time, depending on the sign of $-p+q+\frac{1}{2}$.
\item Case 1 ($\Lambda$ small, $n>1$, initial position $\alpha$): Grain growth is not fast enough to reach the B-mode region. Thus, grains will continue their motion toward the central star in the same stable A-mode.
\item Case 2 ($\Lambda$ large, $n>1$, initial position $\alpha$): Grain growth is very efficient and it reaches the B-mode region at a given time $\Tu$. The growth is then so efficient that the grain will not migrate inside the critical radius $R_{\mathrm{c}}$. Physically, the contribution of the drag integrated over the grain trajectory is too weak to dissipate all the energy and angular momentum of the grain, which is the only way to reach the central region of the disc.
\item Case 3 ($\Lambda \ne 0$, $n<1$, initial position $\beta$): At the beginning of its motion, the growth is efficient enough so that the grain reaches the B-mode region at a time $\Tu$. However, the growth slows down and the grain can not sustain migration in the B-mode. At a time $\Td$, it returns to the A-mode, where it completes it migration toward the central object.
\item Case 4 ($\Lambda$ small, $n>1$, initial position $\gamma$): The acceleration of the grain's growth is not large enough for the B-mode to be stable. The grain reaches the A-mode at a time $\Tu$ and ends its migration in the A-mode.
\item Case 5 ($\Lambda$ large, $n>1$, initial position $\gamma$): The acceleration of the grain's growth is large enough and the B-mode is stable.
\end{itemize}
In this case, we will say that this mode is stable for the grain dynamics.  Mathematically speaking, we have to determine the evolution of $R\left(T\right)$ and $\St\left(T\right)$ from the knowledge of $S(T)$ to describe the grain behaviour.

\subsection{A-mode}

\subsubsection{Grain evolution}

To understand the grain's radial behaviour, we first study the function $R\left(T\right)$ for a grain starting in the A-mode (initial positions $\alpha$ and $\beta$ in Fig. \ref{fig:Modesgrowth}). Starting from Eq. (\ref{eq:Amodgrowth}) and separating variables, a direct integration provides:

\begin{itemize}
\item If $-p+q+\frac{1}{2} \ne 0$:
\begin{equation}
\left\lbrace
\begin{array}{l}
R = \dst\left[1 - \frac{-p+q+\frac{1}{2}}{\Lambda \left(n+1\right)} \left[ \left(S_{0}^{1/n} + \gamma T \right)^{n+1} - S_{0}^{\frac{n+1}{n}}\right] \right]^{\frac{1}{-p+q+\frac{1}{2}}} \\[2ex]
T = \dst\frac{1}{\etaz\Lambda} \left( \left[ \left(1 - R^{-p+q+\frac{1}{2}} \right) \frac{\Lambda \left( n+1 \right)}{-p+q+\frac{1}{2}} + S_{0}^{\frac{n+1}{n}} \right]^{\frac{1}{n+1}} - S_{0}^{1/n}\right) .
\end{array}
\right.
\label{eq:Agrowthu}
\end{equation}

\item If $-p+q+\frac{1}{2} = 0$:
\begin{equation}
\left\lbrace
\begin{array}{l}
R = \dst e^{- \frac{1}{\Lambda \left( n+1 \right)} \left[\left(S_{0}^{1/n} + \gamma T \right)^{n+1} - S_{0}^{\frac{n+1}{n}} \right]} \\
T = \dst\frac{1}{\etaz\Lambda} \left( \left[ - \Lambda \left( n+1 \right) \mathrm{ln}\left(R\right) + S_{0}^{\frac{n+1}{n}}\right]^{\frac{1}{n+1}} - S_{0}^{1/n}\right) .
\end{array}
\right.
\label{eq:Agrowthd}
\end{equation}
\end{itemize}

Thus, in the A-mode, whatever the value of the growth exponent $n$, the grain's behaviour is the direct extension of the results found in absence of growth. If $-p+q+\frac{1}{2} \le 0$, $\lim\limits_{\substack{T \to +\infty}} R = 0$ and the grain never reaches the star. If $-p+q+\frac{1}{2} > 0$, $\lim\limits_{\substack{T \to \Tm}} R = 0$, the grain reaches the star in a finite time
\begin{equation}
\Tm = \frac{1}{\etaz\Lambda} \left( \left[ \frac{\Lambda \left( n+1 \right)}{-p+q+\frac{1}{2}} + S_{0}^{\frac{n+1}{n}} \right]^{\frac{1}{n+1}} - S_{0}^{1/n}\right) .
\label{eq:intermedunA}
\end{equation}

In this case, we study the asymptotic behaviour of $\Tm$ for small and large values of $\Lambda$. For small values of $\Lambda$, we perform a Taylor expansion of $\Tm$ to $\mathcal{O}\left(\Lambda\right)$ and obtain
\begin{equation}
\begin{array}{rl}
\dst \Tm = & \dst\frac{1}{\etaz\sz\left(-p+q+\frac{1}{2}\right)}\left[1 - \frac{n\Lambda}{2\left(-p+q+\frac{1}{2}\right)\sz^{1+\frac{1}{n}}}\right] \\[3ex]
& +\ \mathcal{O}\left(\Lambda^{2}\right) .
\end{array}
\label{eq:intermeddeuxA}
\end{equation}
This expression corresponds to the migration time in the absence of growth \citepalias[see Eq.~24 of][]{Laibe2012} decreased by a term proportional to the growth efficiency. For large values of $\Lambda$,
\begin{equation}
\dst \Tm = \frac{1}{\etaz}\left(\frac{n+1}{-p+q+\frac{1}{2}}\right)^{\frac{1}{\left(n+1\right)}} \Lambda^{-\frac{n}{\left(n+1\right)}} + \mathcal{O}\left(\Lambda^{-\frac{n}{\left(n+1\right)}-1}\right) .
\label{eq:intermedtroisA}
\end{equation}
This expression of $\Tm$ does not depend on $\sz$: the growth is so efficient that it erases the memory of the grain's inital size. However, one has to remember that the expression given by Eq.~\ref{eq:intermedtroisA} is only valid in the A-mode of migration.

\subsubsection{Stability of the A-mode}

To determine if the A-mode is stable or not, we now study the function $\St\left(T\right)$. We study its behaviour near $T = 0$ with a Taylor expansion and in the limit at $T \to \infty$, and then find the maximum value of $\St$ compared to $1$ and determine if at a given time $\Tu$ the grain reaches the B-mode. We use Eqs.~(\ref{eq:Agrowthu}) or (\ref{eq:Agrowthd}) and (\ref{eq:powergrowth}) to calculate $S\left(T\right)$. We first find that in the limit at large time:
\begin{itemize}
\item if $-p+q+\frac{1}{2} > 0$, $\lim\limits_{\substack{T \to \Tm}} \St = 0$,
\item if $-p+q+\frac{1}{2} = 0$,$\lim\limits_{\substack{T \to +\infty}} \St = \lim\limits_{\substack{T \to +\infty}} \mathcal{O}\left(e^{-T^{n+1}}T^{n} \right) = 0$,
\item if $-p+q+\frac{1}{2} < 0$,$\lim\limits_{\substack{T \to +\infty}} \St = \lim\limits_{\substack{T \to +\infty}} \mathcal{O}\left(T^{\frac{\left(n+1\right)p}{-p+q+\frac{1}{2}}+n} \right) = 0$,
\end{itemize}
for $n >0$ and both $p>0$ and $q>0$ (like in protoplanetary discs). Indeed, the exponent of $T$ takes only negative values when $-p+q+\frac{1}{2}<0$. As a result, whatever the surface density and temperature profiles, $\St\left(T\right) \to 0$ at large times. \\

Second, expanding $\St$ at small times provides
\begin{equation}
\begin{array}{rcl}
\St\left(T\right) & = & \sz + \left(S_{0}^{1-\frac{1}{n}} n \gamma - p \etaz \szsq \right)T + \mathcal{O}\left(T^{2}\right)\\
& = & \sz + \etaz n \sz ^{1 - \frac{1}{n}} \left(\Lambda - \Lambdaa \right) T + \mathcal{O}\left(T^{2}\right),
\end{array}
\label{eq:NAexp}
\end{equation}
with
\begin{equation}
\Lambdaa = \frac{p}{n}\sz ^{1+1/n},
\label{eq:defLambdaA}
\end{equation}
and $\St$, from its initial value of $\sz$, increases or decreases with time depending on the sign of $\Lambda-\Lambdaa$.
Thus, if $\Lambda \leq \Lambdaa$, $\St$ initially decreases with respect to time and reaches zero in a finite or infinite time. For $\sz \thicksim 10^{-3}$, $\Lambdaa \thicksim 10^{-3}$, with $\etaz \thicksim 10^{-2}$, $\gamma$ has to be smaller than $10^{-5}$ to be in this regime. On the other hand, if $\Lambda > \Lambdaa$, $\St$ initially increases with time before decreasing to zero, passing through a maximum at $\Tn $. We find $\St\left( \Tn \right) = \mathcal{O}\left(\Lambda ^{\frac{n}{\left(n+1\right)}} \right)$. As a result, there exists a value $\Lambda = \Lambdau$ (necessarily $>\Lambdaa$) such that $ \Lambdau = \mathcal{O}\left(1\right)$ and for which $\St\left( \Tn \right) = \mathcal{O}\left(1\right)$. This is the limiting case for which in a time $\Tu = \Tn$, $\St\left( \Tu \right) = \St\left( \Tn \right) = \mathcal{O}\left(1\right)$.
The different behaviours of growing grains starting their motion in the A-mode are summarized in Fig.~\ref{fig:Amodegrowthfig}. We distinguish three main behaviours:

\begin{figure}
\resizebox{\hsize}{!}{\includegraphics{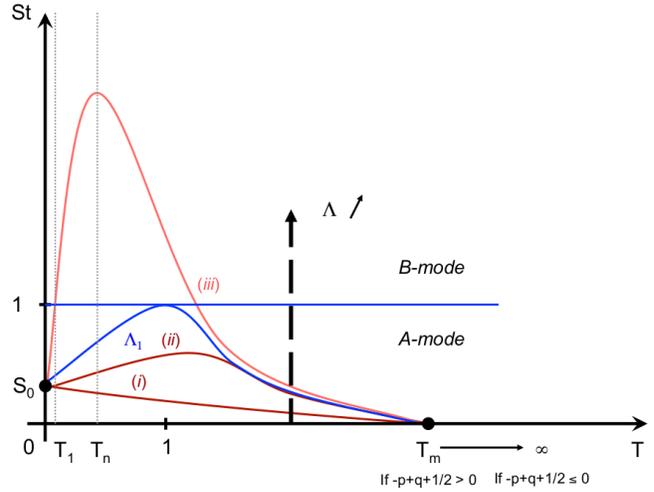}}
\caption{Behaviour of the function $\St\left( T \right)$ for different values of $\Lambda$ for a grain originally starting its migration in the A-mode. See text for details of the three cases. If the growth is efficient enough, grains are able to reach the B-mode of migration.}
\label{fig:Amodegrowthfig}
\end{figure}

\begin{itemize}
\item case ($i$): if $\Lambda \leq \Lambdaa$: $\St\left( T \right) < 1$, the A-mode of migration is stable.
\item case ($ii$): if $\Lambdaa < \Lambda \le \Lambdau$: the grain continues its radial motion in the stable A-mode. If $ \Lambdaa < \Lambda \ll 1$, $\St\left( \Tn \right) \ll 1$. 
\item case ($iii$): if $\Lambda > \Lambdau$: the grain leaves the A-mode of migration for the B-mode. If $ \Lambda \gg 1$ , $\St\left( \Tn \right) \gg 1$, and $\St$ reaches 1 in a time $\Tu \ll \Tn$ for which $\St \ll 1$. 
\end{itemize}

Returning to Fig.~\ref{fig:Modesgrowth}, the behaviour of grains 0 and 1 correspond to cases ($i$) or ($ii$) and the behaviour of grains 2 and 3 to case ($iii$).

\subsection{B-mode}

\subsubsection{Grain evolution}

We now study the radial motion of growing grains in the B-mode of migration (i.e $\St \gg 1$) in the same way. Initially small growing grains can reach the B-mode of migration at a time $\Tu$, where they are located at a radius $\Ru$ and have a size $\Su$ (case ($iii$) of Fig.~\ref{fig:Amodegrowthfig}). They can also start their motion in the B-mode. In this case, $\Su = \sz$, $\Tu = 0$ and $\Ru = 1$. Starting with Eq.~\ref{eq:Bmodegrowth}, a direct integration provides:
\begin{itemize}
\item If $n \ne 1$:
\begin{equation}
\left\lbrace
\begin{array}{rcl}
R & = & \dst\left[\Rua - \frac{p+q+\frac{1}{2}}{\Lambda(1-n)} \left[\left(S_{1}^{1/n} + \gamma T \right)^{1-n}\right.\right.\\[2ex]
&& \qquad\qquad\quad\left.\left.\dst - \left(S_{1}^{1/n} + \gamma \Tu \right)^{1-n}\right] \right]^{\frac{1}{p+q+\frac{1}{2}}} ,\\[2ex]
T & = & \dst\frac{1}{\etaz\Lambda} \left[\left[ \left(\Rua - R^{p+q+\frac{1}{2}} \right) \frac{\Lambda(1-n)}{p+q+\frac{1}{2}}\right.\right.\\[2ex]
&& \qquad\quad\left.\left.\dst + \left(S_{1}^{1/n} + \gamma \Tu \right)^{1-n} \right]^{\frac{1}{1-n}} - S_{1}^{1/n}\right] .
\end{array}
\right.
\label{eq:Bgrowthu}
\end{equation}

\item If $n = 1$:
\begin{equation}
\left\lbrace
\begin{array}{l}
R   =   \dst\left[\Rua - \frac{p+q+\frac{1}{2}}{\Lambda} \,\mathrm{ln}\left(\frac{\Su + \gamma T}{\Su + \gamma \Tu} \right) \right]^{\frac{1}{p+q+\frac{1}{2}}} ,\\[2ex]
T   =  \dst\frac{1}{\etaz\Lambda} \left[\left(\Su + \gamma \Tu \right) e^{\frac{\Lambda \left(\Rua - R^{p+q+\frac{1}{2}} \right)}{p+q+\frac{1}{2}}} -\Su \right] .
\end{array}
\right.
\label{eq:Bgrowthd}
\end{equation}
\end{itemize}

For simplicity, we now consider only the case of $p+q+\frac{1}{2}>0$ (the exponent of $R$ in Eqs.~(\ref{eq:Bgrowthu}) and (\ref{eq:Bgrowthd})), which is the case in real nebulae. Here, contrary to the A-mode, the asymptotic behaviour of $R\left(T\right)$ depends on the value of $n > 0$. If $n \le 1$, $\lim\limits_{\substack{T \to \Tm}} R = 0$ (weak growth). If $n > 1$, two cases arise. First, if $\Lambda \le \dst\frac{\left(p+q+\frac{1}{2}\right) \left(S_{1}^{1/n} + \gamma \Tu \right)^{1-n}}{\Rua \left(n-1\right)} = \Lambda_{\mathrm{c}}$, \mbox{$\lim\limits_{\substack{T \to \Tm}} R = 0$} (weak growth). Second, if $\Lambda > \Lambda_{\mathrm{c}}$, $\lim\limits_{\substack{T \to +\infty}} R = \Rc$, the grain asymptotically approaches a radius
\begin{equation}
\Rc = \Ru \dst\left(1 - \frac{\left(p+q+\frac{1}{2}\right) \left(S_{1}^{1/n} + \gamma \Tu \right)^{1-n}}{\Rua \left(n-1\right) \Lambda} \right)^{\frac{1}{p+q+\frac{1}{2}}}
\label{eq:eqn-Rc}
\end{equation}
(intense growth).

\subsubsection{Stability of the B-mode}

To determine in which case the B-mode is stable, we study the asymptotic behaviour of $\St\left(T\right)$ for small and large $\Lambda$. 
\begin{figure}
\resizebox{\hsize}{!}{\includegraphics{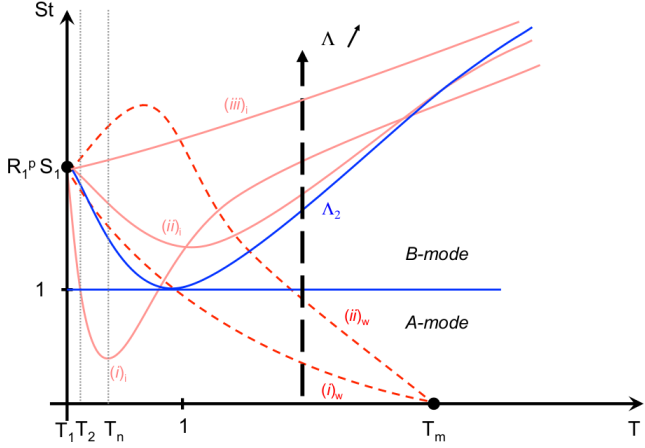}}
\caption{Behaviour of the function $\St\left( T \right)$ for different values of $\Lambda$ for a grain originally starting its migration in the B-mode. The five behaviours are described in the text. An efficient growth is required to maintain the particle in the B-mode.}
\label{fig:Bmodegrowthfig}
\end{figure}
First, in the limit at large time:
\begin{itemize}
\item For weak growth (if $n \le 1$ or $n > 1$ and $\Lambda\le\Lambda_{\mathrm{c}}$): $\lim\limits_{\substack{T \to \Tm}} \St = 0$. Consequently, whatever the initial evolution of $\St$, there is a time $T_{2}$ for which the grain switches to the A-mode: the B-mode of migration is unstable. This situation corresponds to trajectories~3 and 4 of Fig.~\ref{fig:Modesgrowth}.
\item For strong growth (if $n > 1$ and $\Lambda > \Lambda_{\mathrm{c}}$): $\lim\limits_{\substack{T \to +\infty}} \St = +\infty$. In this case, the growth is efficient, like for trajectory~5 of Fig.~\ref{fig:Modesgrowth}: the grain can not drift inward past a critical radius $\Rc$.
\end{itemize}

Second, expanding $\St$ at times close to $\Tu$ provides
\begin{equation}
\begin{array}{rcl}
\St & = & \dst R_{1}^p\left(S_{1}^{1/n} + \gamma \Tu \right)^{n} +  R_{1}^p \left(S_{1}^{1/n} + \gamma \Tu \right)^{n-1} \times \\[1ex]
&& \dst\left(n \gamma - \frac{\gamma p}{\Lambda \Rua}\left(S_{1}^{1/n} + \gamma \Tu \right)^{-n+1} \right)(T-\Tu) \\[1ex]
&& \dst + \mathcal{O}\left((T-\Tu)^{2}\right) \\[2ex]
& = & \dst R_{1}^p\left(S_{1}^{1/n} + \gamma \Tu \right)^{n} + R_{1}^p\left(S_{1}^{1/n} + \gamma \Tu \right)^{n-1} \times \\[1ex]
&& \dst n \etaz(\Lambda-\Lambdab)(T-\Tu) + \mathcal{O}\left(T-\Tu)^{2}\right) ,
\label{eq:expandN}
\end{array}
\end{equation}
where:
\begin{equation}
\Lambdab = \frac{p}{n} \frac{\left(S_{1}^{1/n} + \gamma \Tu \right)^{1-n}}{\Rua} ,
\label{eq:lambdabdef}
\end{equation}
and $\St$, from its value of $\Su$ at $\Tu$, increases or decreases with time depending on the sign of $\Lambda-\Lambdab$.
\begin{figure*}
\centering
\includegraphics[width=17cm]{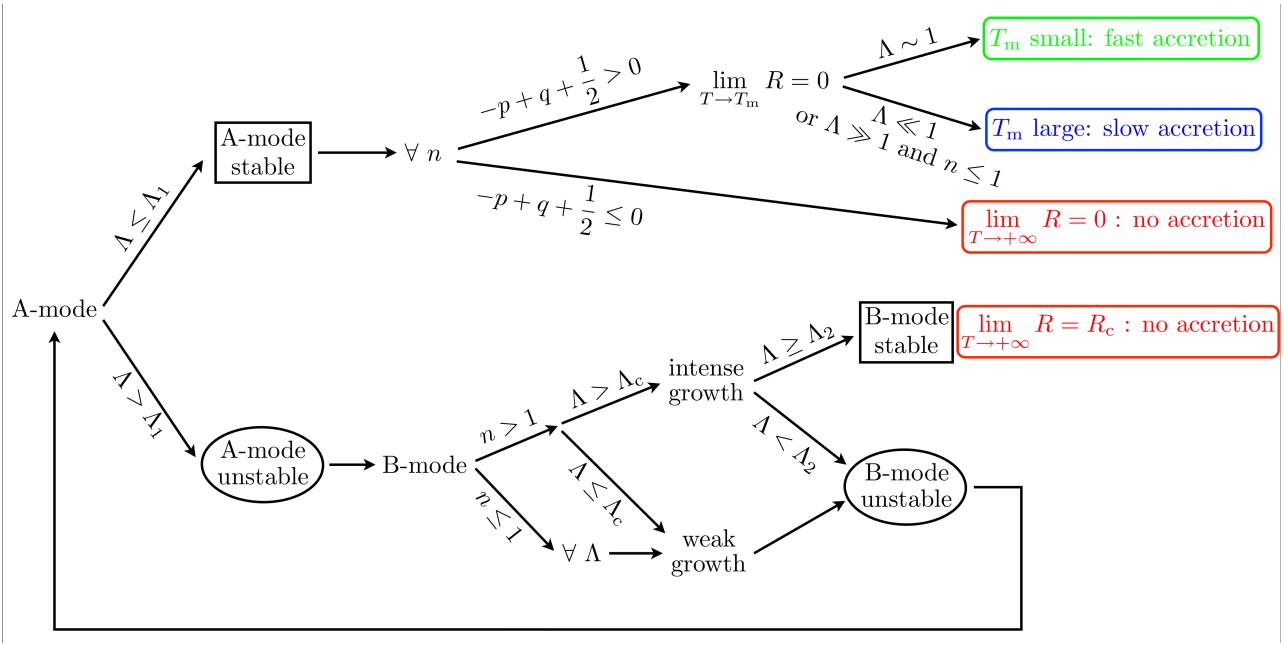}
\caption{Summary of the global evolution of an initially small growing grain. For intense accelerating growth, particles decouple from the gas at a finite radius, staying in the B-mode of migration. Otherwise, grains survive the radial-drift barrier if they pile-up, i.e. if $-p+q+1/2\le 0$ since they migrate in the A-mode. The grains radial migration is always more efficient when the ratio between growth and migration timescales is of order unity.}
\label{fig:ModesCycle}
\end{figure*}
Thus, if $\lim\limits_{\substack{T \to \Tm}} \St = 0 $ and $\Lambda < \Lambdab$, $\St\left(T\right)$ is initially decreasing and tends to zero at large times. If $\lim\limits_{\substack{T \to \Tm}} \St = 0 $ and $\Lambda > \Lambdab$, $\St$ is initially increasing, reaches a maximum value and tends to zero at large times. If $\lim\limits_{\substack{T \to +\infty}} \St = +\infty $ and $\Lambda < \Lambdab$, $\St$ is initially decreasing, reaches a minimum value at a time $\Tn$ and tends to infinity at large times. There exists a value $\Lambda = \Lambdad$ (necessarily $<\Lambdab$), for which $\St\left(\Tn\right) =1$. Thus, if $\Lambda < \Lambdad$, $\St\left(\Tn\right) < 1$ and if $\Lambda > \Lambdad$, $\St\left(\Tn\right) > 1$. If $\lim\limits_{\substack{T \to +\infty}} \St = +\infty $ and $\Lambda > \Lambdab$, $\St\left(T\right)$ is initially increasing and tends to infinity at large times.

The different behaviours of growing grains experiencing a radial motion in the B-mode are summarized in Fig.~\ref{fig:Bmodegrowthfig}. We distinguish five main behaviours:

\begin{itemize}
\item case $(i)_\mathrm{w}$: weak growth and $\Lambda < \Lambdab$: $\St$ decreases and the B-mode is unstable.
\item case $(ii)_{\mathrm{w}}$: weak growth and $\Lambda \ge \Lambdab$: $\St$ increases and then decreases as it tends to zero, the B-mode is thus unstable. Therefore, whatever the value of $\Lambda$, the grain will always reach the A-mode. 
\item case $(i)_{\mathrm{i}}$: intense growth and $\Lambda <\Lambdad$: Even though $\St$ reaches values large enough at large time, its initial decrease due to small values of $\Lambda$ leads to a transition to the A-mode. 
\item case $(ii)_{\mathrm{i}}$: intense growth and $\Lambdad \le \Lambda < \Lambdab$:  Even if $\St$ decreases at small times, $\Lambda$ is large enough for the grain to stay in the B-mode.
\item case $(iii)_{\mathrm{i}}$: intense growth and $\Lambda \ge \Lambdab$: $\St$ increases and the B-mode is stable.

\end{itemize}

Returning to Fig. \ref{fig:Modesgrowth}, the behaviour of grains~3 and 4 corresponds to cases~$(i)_\mathrm{w}$ and $(ii)_\mathrm{w}$, for which the B-mode is unstable, and the behaviour of grains~2 and 5 corresponds to cases~$(ii)_\mathrm{i}$ and $(iii)_\mathrm{i}$, for which the B-mode is stable.

\subsection{Combining A- and B-modes}

The radial behaviour of initially small growing grains for the different values of the parameters can now be interpreted and a summary is presented in Fig.~\ref{fig:ModesCycle}. If $\Lambda \le \Lambdaa$, or if $\Lambdaa < \Lambda \le \Lambdau = \mathcal{O}\left(1\right)$, the growth is not efficient enough to reach the B-mode of migration and the A-mode is stable. The dust behaviour at large times depends on the value of $-p+q+\frac{1}{2}$ (as for non-growing grains, which ultimately migrate in the A-mode). If $-p+q+\frac{1}{2}>0$, the radial drift efficiency is increased by growth and the particle is accreted in a finite time which decreases with $\Lambda$. If $-p+q+\frac{1}{2} \le 0$ the particle reaches the inner region of the disc, but will never be accreted onto the central object.

 If $\Lambda > \Lambdau$, the growth is efficient enough and the A-mode is unstable. The grain reaches the B-mode at a rate that depends on  $\Lambda$. For weak growth, or for intense growth with $\Lambda <\Lambdad$, the B-mode is unstable and the particle reaches the A-mode again.  However, the new value of  $\Lambdau$ which determines the stability of this migration in the A-mode depends of the new initial position in the disc. If $\Lambdau$ remains small enough, an oscillation between the two modes may occur  until $\St$ is large enough. Then, the A-mode will be stable as $\Lambdau$ reaches large enough values. Again, the final grain behaviour depends on $-p+q+\frac{1}{2}$. For intense growth with either $\Lambdad \le \Lambda < \Lambdab$ or $\Lambda \ge \Lambdab$, the growth is very efficient. The B-mode is stable and the particles converge to a limit radius $\Rc$ in an infinite time. In this case, the fact that the particle is not accreted onto the central star does not depend on the profile effect \citepalias[described in][and mentioned in Sect.~\ref{sec:nut}]{Laibe2012} as for a weak growth process, but on the ``growth effect", which is defined by the value of $\Lambda$.
 
Finally, the results of the analysis we performed analytically in this section are confirmed by a direct integration of the equations of motion in Appendix~\ref{app:complements}.

\section{Conclusions and perspectives}
\label{sec:conclusion}

We have studied the impact of grain growth onto the radial motion of grains in protoplanetary discs using two  toy models: a linear and a power law of grain growth. We found that:

\begin{enumerate}

\item Depending on  the relative efficiency between growth and migration which is given by the value of the dimensionless parameter $\Lambda$, three behaviours are possible for the radial motion of the dust grains. If $\Lambda \ll 1$, the impact of the growth on the migration consists only of a slowly varying correction compared to the case without any growth. Conversely, for $\Lambda \gg 1$, the migration is efficiently slowed down by the growth. Indeed, grains spend a relatively little time in the regime where $\St = \mathcal{O}(1)$ and become more and more decoupled from  the gas as they grow. Thus, the migration occurs on a timescale much longer than that without any growth since the time where the particle stays in the regime $\St =\mathcal{O}(1)$ is negligible. For the intermediate case --- i.e. $\St =\mathcal{O}(1)$--- the growth and the migration are occurring together, leading to efficient radial drift.

\item The final outcome of the grain's radial migration is however a result of its global motion through the disc. The grains experience a pile-up from the competition between the drag and the increasing pressure gradient if $-p+q+1/2\le0$. If the grain does not experience any pile-up, there exists a value of $\Lambda$ such that $\Lambda = \mathcal{O}\left(1 \right)$ --- i.e. a critical value for migration $\gamma =  \mathcal{O}\left(\etaz \right)$--- which minimises the time for the grain to reach the central star.

\item In addition to the standard grain pile up close to $R=0$ which occurs for non-growing grains, the discs can retain solid material with a high growth efficiency. Indeed, for $n>1$, the particle can end its migration in the B-mode and converges in an infinite time to a finite radius $\Rc$. This growth effect happens when the grain size increases so quickly that the integrated contribution of the drag can not dissipate either the grain energy or its angular momentum.

\item Allowing the growth to accelerate or decelerate enriches the grains possible behaviours. As shown in Fig.~\ref{fig:ModesCycle}, we distinguish four regimes for the radial motion of growing dust grains: (1) accretion enhanced by growth, (2) accretion slowed down by growth, (3) no accretion by traditional pile up and (4) no accretion by decoupling at a finite radius, depending on sign of $-p+q+\frac{1}{2}$, and the values of $\Lambda$ and $n$.

\end{enumerate}

While useful, these toy growth models are not sufficient to quantitatively predict the behaviour of the grains in protoplanetary discs since the growth rate of the particles does not depend on the radial position of the grains. This assumption is incorrect for physical growth models. In this case, the growth and the radial drift of the grain interact and the grains dynamics may be complicated. This issue is addressed in Paper~II.

\section*{acknowledgements}
This research was partially supported by the Programme National de Physique
Stellaire and the Programme National de Plan\'etologie of CNRS/INSU, France,
and the Agence Nationale de la Recherche (ANR) of France through contract
ANR-07-BLAN-0221. STM acknowledges the support of a Swinburne Special Studies Program. GL is grateful to the Australian Research Council for funding via Discovery project grant DP1094585. JFG's research was conducted within the Lyon Institute of Origins under grant ANR-10-LABX-66. The authors want to thank D. Price for useful comments. 

\bibliography{bibliodelta}

\begin{appendix}
\section{Notations}
\label{App:Notations}

The notations and conventions used throughout this paper are summarized in Table~\ref{tabnote}.
\begin{table}
\begin{center}
\begin{tabular}{ll}
\hline Symbol & Meaning \\ \hline
$M$ & Mass of the central star \\
$\textbf{g}$ & Gravity field of the central star \\ 
$\Rz$ & Initial distance to the central star \\
$\rhog$ & Gas density \\
$\brhog\left(r\right)$ & $\rhog\left(r,z=0\right)$ \\
$\cs$ & Gas sound speed \\
$\bcs\left(r\right)$ & $\cs\left(r,z=0\right)$ \\
$\csz$ & Gas sound speed at $\Rz$ \\
$T$ & Dimensionless time\\
$\mathcal{T}$ & Gas temperature ($\mathcal{T}_{0}$: value at $\Rz$)\\ 
$\Sigmaz$ & Gas surface density at $\Rz$ \\
$p$ & Radial surface density exponent \\
$q$ & Radial temperature exponent \\
$P$ & Gas pressure \\
$v_{\mathrm{k}}$ & Keplerian velocity at $r$ \\
$\vkz$ & Keplerian velocity at $\Rz$ \\
$\Hz$ & Gas scale height at $\Rz$ \\
$\phiz$ & Square of the aspect ratio $\Hz/\Rz$ at $\Rz$ \\
$\etaz$ & Sub-Keplerian parameter at $\Rz$ \\
$s$ & Grain size \\
$S$ & Dimensionless grain size \\
$\sz$ & Initial dimensionless grain size \\
$\St$& Stokes number \\
$y$ & Grain size exponent in the drag force  \\
$\textbf{v}_{\mathrm{g}}$ & Gas velocity \\
$\textbf{v}$ & Grain velocity \\
$\rhod$ & Dust intrinsic density \\
$\md$ & Mass of a dust grain \\
$\ts$ & Drag stopping time \\
$\tsz$ & Drag stopping time at $\Rz$ \\
\hline
\end{tabular}
\end{center}
\caption{Notations used in the article.}
\label{tabnote}
\end{table}
%

\section{Radial evolution with the power-law toy model}
\label{app:complements}

To compare this analysis for the evolution directly obtained from the equation of motion, we integrate numerically Eq.~\ref{eq:eqradwithgrowth} for different values of the parameters $\etaz$, $\sz$, $p$, $q$, $\gamma$ and $n$. We set the dimensionless acceleration due to the pressure gradient $\etaz = 10^{-2}$ to mimic a real nebula and an initial size $\sz= 10^{-3}$ to start with grains small enough (like in the planet formation process) so as to initially have $\St \ll 1$. We verify that with $\sz = 10^{-4}$ or $10^{-2}$ the results do not significantly change. We then explore the parameter space by first varying the order of magnitude of $\gamma$ from $10^{-4}$ to $10^{2}$ and second setting $n = 0.5 ,1 ,2$ so as to treat one convex, one linear and one concave evolution of $\St(T)$. Thus, $7 \times 3 = 21$ trajectories are computed for each of the nine ($p$ , $q$) pairs formed by $p = 0 , 3/4 , 3/2$ and $q = 1/2 , 3/4 ,1$.

\begin{figure}
\begin{center}
\resizebox{\hsize}{!}{\includegraphics[angle=-90]{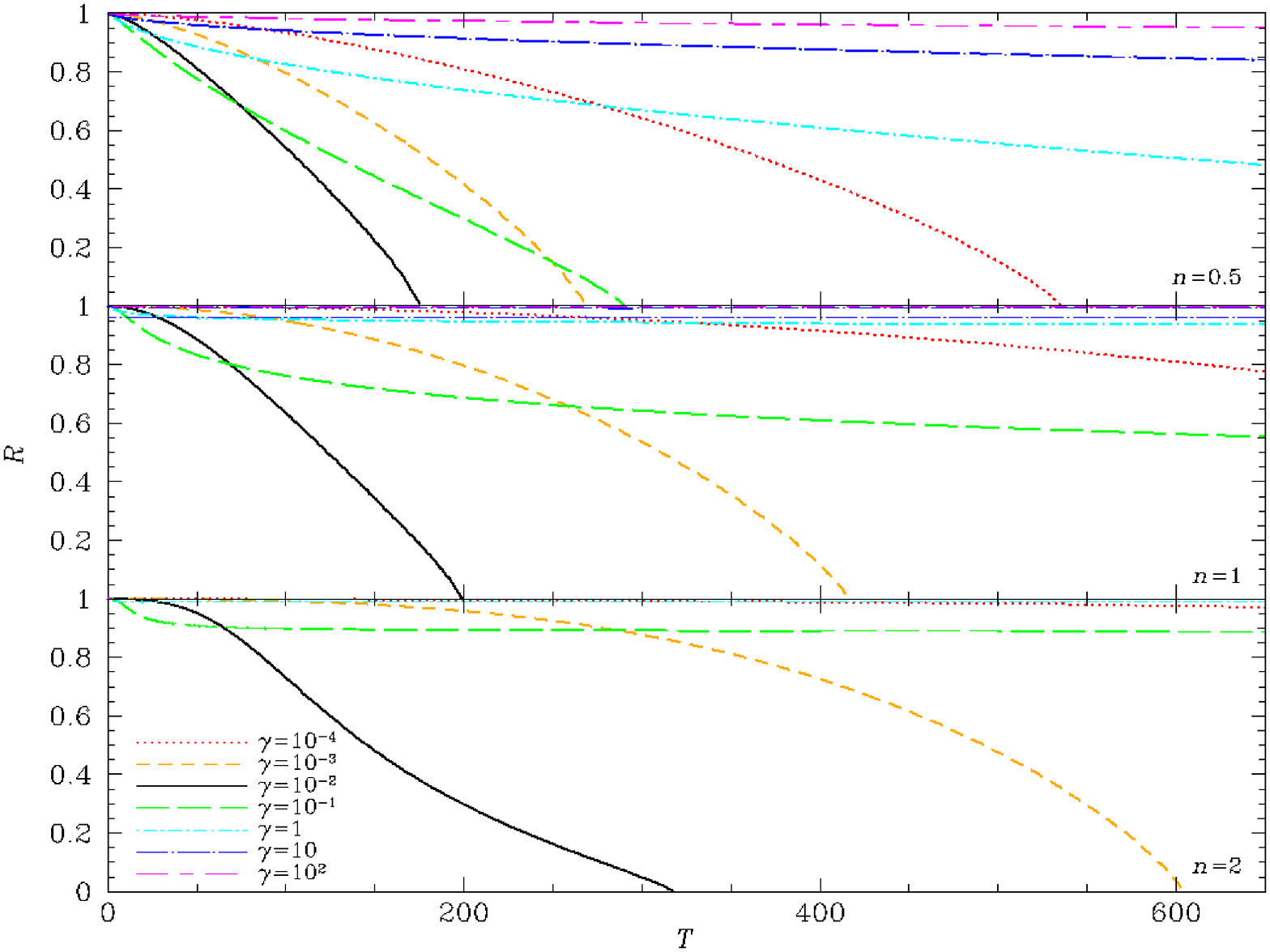}}
\resizebox{\hsize}{!}{\includegraphics[angle=-90]{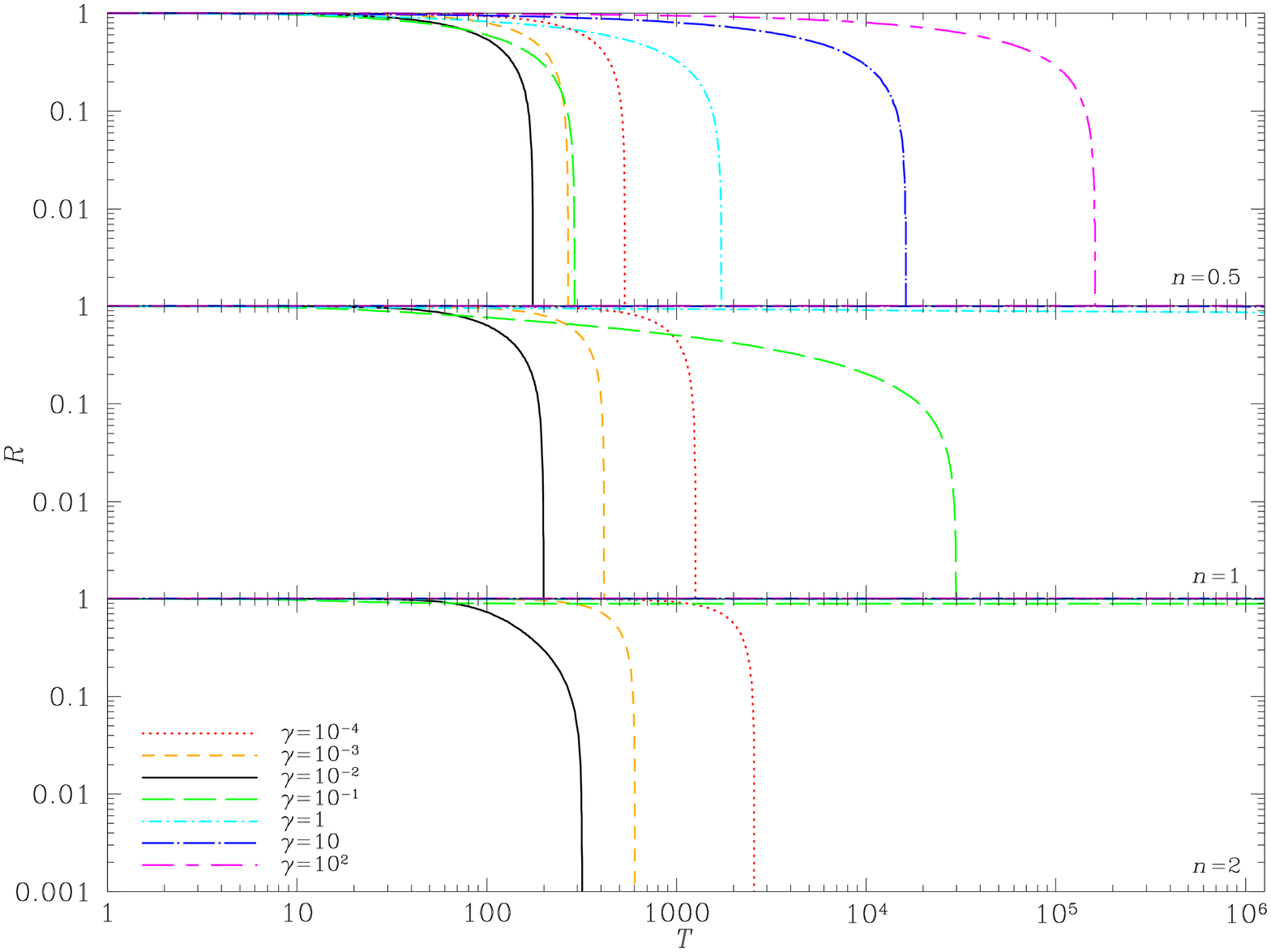}}
\caption{Radial motion $R\left(T\right)$ of dust grains for $\etaz = 10^{-2}$, $\sz = 10^{-3}$, $p=0$, $q=3/4$, and values of $\gamma$ ranging from $10^{-4}$ to $10^2$ in linear (top panel) and logarithmic scale (bottom panel). From top to bottom in each panel: $n = 0.5$, 1, and 2.} 
\label{fig:plotfirstcasegrowth}
\end{center}
\end{figure}

\begin{figure}
\begin{center}
\resizebox{\hsize}{!}{\includegraphics[angle=-90]{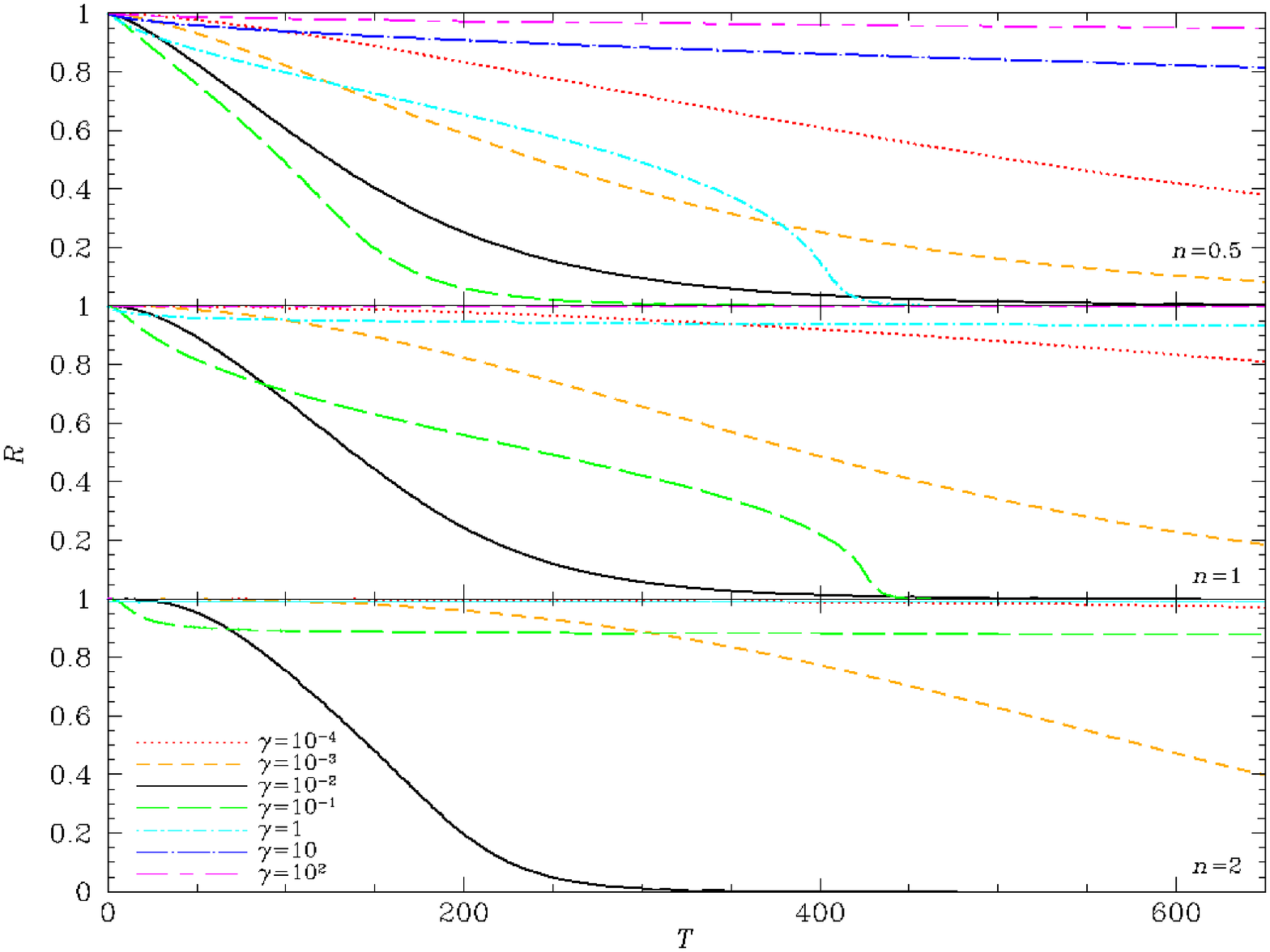}}
\resizebox{\hsize}{!}{\includegraphics[angle=-90]{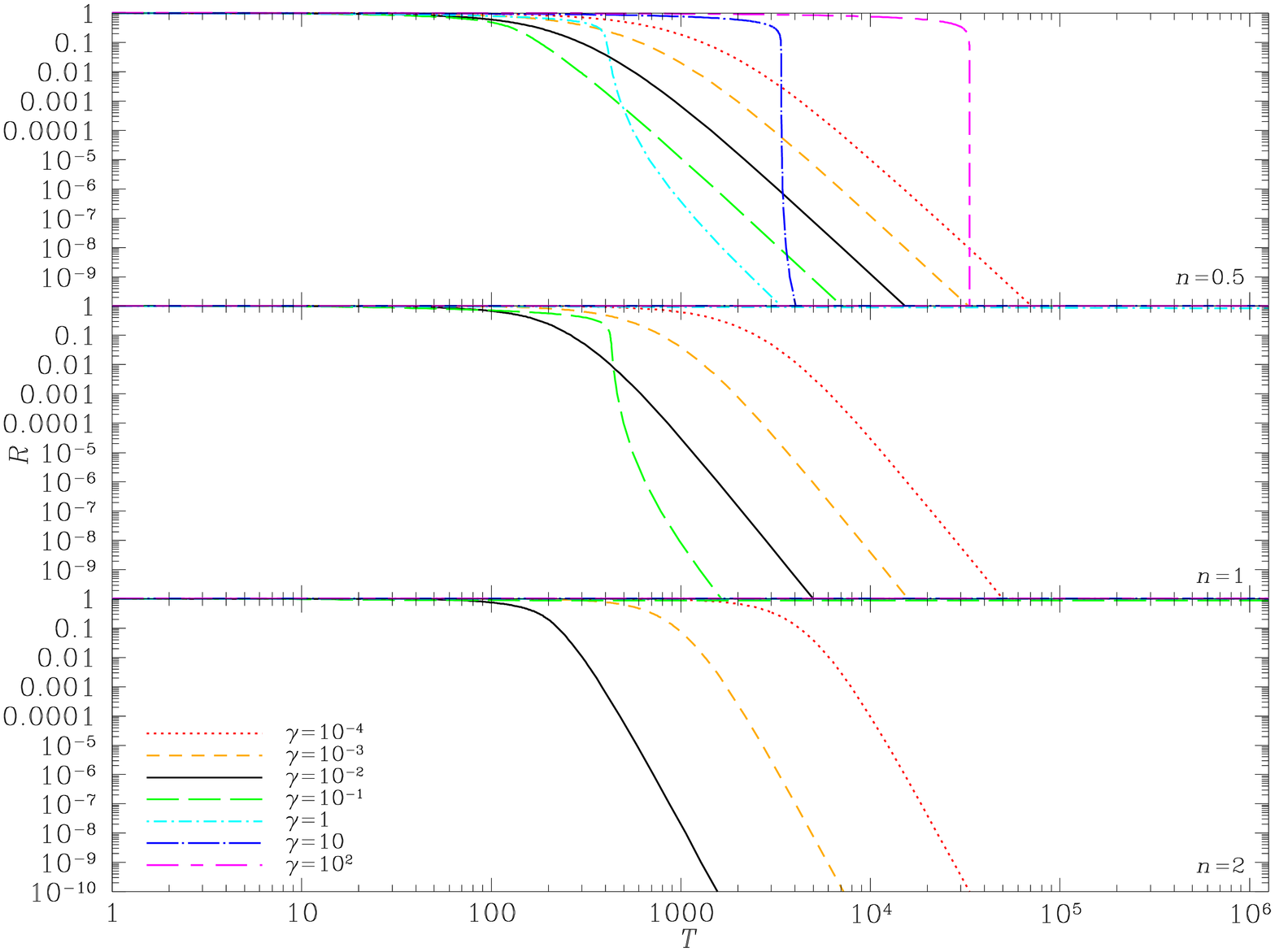}}
\caption{Same as Fig.~\ref{fig:plotfirstcasegrowth} for $p=3/2$ and $q=3/4$.} 
\label{fig:plotsecondcasegrowth}
\end{center}
\end{figure}

To show how growth affects the radial grain motion, we focus on the two cases: ($p=0$, $q=3/4$) and ($p=3/2$, $q=3/4$), shown in Figs.~\ref{fig:plotfirstcasegrowth} and \ref{fig:plotsecondcasegrowth}, which we analysed in absence of grain size evolution in \citetalias{Laibe2012}.
We note that a large class of migration regimes occur, depending on the growth parameters $\gamma$ and $n$. The following major behaviours can be highlighted:

\begin{itemize}

\item When varying $\gamma$ while fixing all the other parameters, migration to the inner regions is optimal for a value which depends on $p$, $q$ and $n$ and is close to $10^{-2}$. With $\etaz = 10^{-2}$, this corresponds to $\Lambda = 1$. The corresponding migration time for $R$ to reach 0 is of the order of $10^{2} = \etaz ^{-1}$, which agrees with the result found above (see Fig.~\ref{fig:plotKummer}, which is only valid for $n=1$, or Eq.~\ref{eq:intermedunA}). When $\gamma \ge 1$, the grain migrates slowly to the central star or to the limit radius, $R_{\mathrm c}$, as predicted by Eq.~\ref{eq:eqn-Rc}.

\item When increasing $n$ while fixing all the other parameters, migration at small and large values of $\gamma$ is less efficient. 

\item When comparing the case $p = 3/2$ ($-p+q+\frac{1}{2}<0$) and the case $p = 0$ ($-p+q+\frac{1}{2}>0$) while fixing all the other parameters, the dust dynamics is modified by the grains pile-up. For example, comparing the cases at $\gamma = 10^{-2}$, even with different values of $n$ for each value of $p$, the particle is accreted onto the central star in a finite time for $-p+q+\frac{1}{2}>0$, but only migrates slowly to the disc inner regions for $-p+q+\frac{1}{2}<0$ (the curvature of the trajectory becomes convex at large time). 

\end{itemize}

Thus, the results obtained from the equation of motion confirm the prediction performed from (1) the small pressure gradient and (2) the A- and B-modes expansions. It is also important to note that the grain's radial trajectories are given in dimensionless coordinates. In some cases (for example $p=3/2$, $q=3/4$, $\etaz = 10^{-2}$, $\gamma = 10^{-2}$, see the black curve in Fig.~\ref{fig:plotsecondcasegrowth}), the particle drastically slows down between $R = 0.05$ for $n = 0.5$ and $R = 0.1$ for $n = 2$ because of the profile effect. This corresponds to 10~AU for a grain which starts to migrate at 200~AU, which is still relevant for planet formation. Moreover, even if the accretion time onto the central star is finite, the accretion process can be long enough, especially for large values of $\Lambda$ or values of $-p+q+\frac{1}{2}$ close to zero to allow particles to grow in real nebulae before the disc disappears.

\end{appendix}

\end{document}